\documentclass[11pt]{article}
\usepackage{moriond,epsfig}
\usepackage{amsmath}
\usepackage{amssymb}
\usepackage{amsfonts}
\usepackage{mathrsfs}
\usepackage{graphicx}

\newcommand{\raw}{\rightarrow}

\newcommand{\be}{\begin{equation}}
\newcommand{\ee}{\end{equation}}
\newcommand{\bea}{\begin{eqnarray}}
\newcommand{\eea}{\end{eqnarray}}

\def\be{\begin{equation}}
\def\ee{\end{equation}}
\def\bea{\begin{eqnarray}}
\def\eea{\end{eqnarray}}

\begin{document}
\vspace*{4cm}
\title{ CP ASYMMETRIES FROM NON-UNITARY LEPTONIC MIXING}

\author{ J. L\'OPEZ-PAV\'ON }

\address{Departamento de F\'\i sica Te\'orica and Instituto de F\'\i sica Te\'orica UAM/CSIC,
\\
Universidad Aut\'onoma de Madrid, 28049 Cantoblanco, Madrid, Spain}

\maketitle\abstracts{ A low-energy non-unitary leptonic mixing matrix is a generic effect of some theories of new physics accounting for neutrino masses. We show how the extra CP-phases of a general non-unitary matrix allow for sizeable CP-asymmetries in the $\nu_\mu\rightarrow \nu_\tau$ channel. This CP-asymmetries turns out to be an excellent probe of such new physics.}. %In adition, we clarify the relationship betweeen our framework and the so called "non-standard neutrino interactions" scenarios: the sensitivities explored here apply as well to such constructions, except for extremely fine-tuned cancellations.

\section{Introduction}
\label{intro}
Non-zero neutrino masses constitutes one of the main evidence for physics beyond the Standard Model of particle physics (SM). In the complete theory accounting for them and encompassing the SM the complete mixing matrices should be unitary, as mandated by probability conservation. However, the effective $3 \times 3$ submatrices describing the mixing of the light fermionic fields need not to be unitary, since these known fields in the theory may mix with other degrees of freedom. %We will not particularize to any given model of new physics beyond the SM (BSM) and 
We will assume that the full theory is indeed unitary, whereas, low-energy non-unitarity may result from BSM physics contributing to neutrino propagation, when the physical measurements are described solely in terms of SM fields~\cite{uni}. Specifically, the tree-level exchange of heavy fermions (scalars) will (not) induce low-energy non-unitary contributions through dimension six effective operators~\cite{Abada:2007ux}.

In Ref.~\cite{uni} the so-called MUV (minimal unitarity violation) was developed and the {\it absolute} values of the elements of the matrix N were determined. %, using data from neutrino oscillation experiments and weak decays. 
However, no information on the phases of the mixing matrix is available, neither on the standard phases nor on the new non-unitary ones, as present oscillation data correspond mainly to disappearance experiments. Here we will explore the future sensitivity to the new CP-odd phases of the leptonic mixing matrix associated to non-unitary. In particular, it will be shown that CP-asymmetries in the $\nu_\mu\raw\nu_\tau$ channel are an excellent probe of such new physics. Notice that CP-odd effects in that channel are negligible in the standard unitary case, in which the golden channel for CP-violation is $\nu_e\raw\nu_\mu$.

\section{Formalism}
\label{formalism}

We start from the parametrization of the general non-unitary matrix $N$, which relates flavor and mass fields, as the product of an hermitian and a unitary matrix, defined by
\be
\label{N}
\nu_{\alpha} = N_{\alpha i}\, \nu_{i}\equiv\left[ (1+\eta) U\right]_{\alpha i}\nu_{i}   \, ,
\ee
with $\eta^\dagger=\eta$. Since $NN^\dagger=(1+\eta)^2\approx 1+2\eta$, the bounds derived in Ref.~\cite{uni} for the modulus of the elements of $NN^\dagger$ can also be translated into constraints on the elements of $\eta$. It follows that
\medskip
\begin{eqnarray}
|\eta| =
\begin{pmatrix}
 <5.5\cdot 10^{-3}  &  < 3.5 \cdot 10^{-5}  &    < 8.0 \cdot 10^{-3} \\
 < 3.5 \cdot 10^{-5}  & <5.0\cdot 10^{-3}  &    < 5.1 \cdot 10^{-3} \\
 < 8.0 \cdot 10^{-3} &  < 5.1 \cdot 10^{-3}
&  <5.0\cdot 10^{-3}
\end{pmatrix},
\label{limits}
\end{eqnarray}
at the $90\%$ confidence level. The bounds %on $\eta_{\mu \tau}$ 
have been updated with the latest experimental bound on $\tau \rightarrow \mu\gamma$~\cite{taumugamma}. Eq.~(\ref{limits}) shows that the matrix $N$ is constrained to be unitary at the per cent level accuracy or better. Therefore, the unitary matrix $U$ in Eq.~(\ref{N}) can be identified with the usual unitary mixing matrix $U=U_{PMNS}$, within the same accuracy. The flavor
eigenstates can then be conveniently expressed as \footnote{The handy superscript $SM$ is an abuse of language, to describe the flavor eigenstates of the standard unitary analysis.}
\be
|\nu_\alpha> =\dfrac{(1+\eta^*)_{\alpha \beta}U^*_{\beta i}}{\left[ 1+2\eta_{\alpha\alpha}+(\eta^2)_{\alpha\alpha}\right]^{1/2}}\,|\nu_i>\equiv
\dfrac{(1+\eta^*)_{\alpha \beta}}{\left[ 1+2\eta_{\alpha\alpha}+(\eta^2)_{\alpha\alpha}\right]^{1/2} }\,|\nu^{SM}_\beta> .
\label{estados}
\ee
So the neutrino oscillation amplitude, neglecting terms quadratic in $\eta$, is given simply by
\be
<\nu_\beta|\nu_\alpha (L)> =  A^{SM}_{\alpha \beta}(L)\,\left(1-\eta_{\alpha\alpha}-\eta_{\beta\beta} \right) +%\nonumber\\
\sum_\gamma \left( \eta^*_{\alpha \gamma}A^{SM}_{\gamma \beta}(L)+
\eta_{\beta \gamma}A^{SM}_{\alpha \gamma}(L) \right)\,,
\label{amplitud}
\ee
\noindent
with
\be
A^{SM}_{\alpha \beta}(L)\equiv<\nu^{SM}_\beta|\nu^{SM}_\alpha (L)>
\ee
\noindent
being the usual oscillation amplitude of the unitary analysis.

In order to study the posible new CP-violation signals arising from the new phases in $\eta$, one has to consider appearance channels, $\alpha\ne\beta$.  The best sensitivities to such phases will be achieved in a regime where the first term in Eq.~(\ref{amplitud}) is suppressed. This is possible at short enough baselines, where the standard appearance amplitudes are very small while $A^{SM}_{\alpha \alpha}(L)\simeq 1$. In that case the total amplitude is well approached by
\be
<\nu_\beta|\nu_\alpha (L)> =  A^{SM}_{\alpha \beta}(L)+ 2\eta_{\alpha \beta}^* + \mathcal{O}(\eta\,A),
\label{amp}
\ee
where $\mathcal{O}(\eta\,A)$ only includes appearance amplitudes and $\eta\,$ components with flavor indices other than $\alpha \beta$. Then, at short enough baselines, each oscillation probability in a given flavor channel, $P_{\alpha \beta}$, is mostly sensitive to the corresponding
$\eta_{\alpha \beta}$. The rest of the elements of the $\eta$ matrix can be safely disregarded in the analysis below, without implying to assume zero values for them. That is, their effect is subdominant, a fact that has been numerically checked for the main contributions. %This also means that the subleading corrections from the cross sections and fluxes discussed in Ref.~\cite{uni}, which induce also $\mathcal{O}(\eta\,A)$ corrections, do not need to be taken into account.

For instance, within the above-described approximation, the oscillation probability for two families would read:
\be
P_{\alpha \beta} = \sin^{2}(2\theta)\sin^2\left(\frac{\Delta L}{2}\right)%\nonumber
%\ee
%\be
-4|\eta_{\alpha \beta}|\sin\delta_{\alpha \beta}\sin(2\theta)\sin\left(\Delta L\right)
+4|\eta_{\alpha \beta}|^2,
\label{prob}
\ee
where $\Delta = \Delta m^2 /2E$ and $\eta_{\alpha \beta}=|\eta_{\alpha \beta}|e^{-i\delta_{\alpha \beta}}$.
The first term in the above equation is the standard oscillation probability. The third term is associated to the zero-distance effect stemming from the non-orthogonality of the flavor eigenstates~\cite{uni}. Finally, the second term is the CP-violating interference between the other two which we are interested in. Notice that, even in two families, there is CP-violation due to the non-unitarity.

\section{Sensitivity to the new CP-odd phases: the $\nu_\mu \raw \nu_\tau$ channel}
\label{sensitivity}

This channel is the best option to study the CP-violation comming from non-unitarity. This is because present constraints on $\eta_{e \mu}$ are too strong to allow a signal in the $\nu_e \rightarrow \nu_\mu$ one (see Eq.~(\ref{limits})), and $\nu_e \rightarrow \nu_\tau$ has extra supressions by small standard parameters such as $\sin\theta_{13}$ or $\Delta_{12}$~\cite{paper}. In the numerical computation of $P_{\mu \tau}$, the only approximation performed is to neglect all $\eta$ elements but $\eta_{\mu \tau}$. They should be indeed subdominant (see Eq.~(\ref{amp})). This approximation has been checked numerically.

Eq.~(\ref{prob}) suggests us that the best sensitivities to CP-violation will be achieved at short baselines and high energies, where the standard term is suppressed by $\sin^2(\frac{\Delta L}{2})$. Therefore, we will study a Neutrino Factory beam~\cite{nf} resulting from the decay of $50$ GeV muons, to be detected at a $130$ Km baseline, which matches for example the CERN-Frejus distance. For this set-up, $\sin(\frac{\Delta_{31} L}{2})\simeq 1.7\cdot10^{-2}$ and $\sin(\frac{\Delta_{21} L}{2})\simeq 6\cdot10^{-4}$, where $\Delta_{jk}\equiv(m^2_j-m^2_k)/2E$. Then, if the $\eta_{\mu\tau}$ value is close to their experimental limit in Eq.~(\ref{limits}), all terms in the  oscillation probability given in  Eq.~(\ref{prob}) can be of similar order for the channel $\nu_\mu \raw \nu_\tau$. On the experimental side, we will assume $2\cdot10^{20}$ useful decays per year and five years running with each polarity, consider a 5 Kt Opera-like detector and, finally, sensitivities and backgrounds a factor 5 larger~\cite{paper} than those used for the $\nu_e\rightarrow\nu_\tau$ channel in Ref.~\cite{silver}.

Since this channel is not suppressed by small standard parameters such as $\sin\theta_{13}$ or $\Delta_{12}$, the two family approximation in Eq.~(\ref{prob}), with $\theta=\theta_{23}$ and $\Delta=\Delta_{31}$, is very accurate to understand the results \footnote{The complete expanded expression for $P_{\mu\tau}$ can be found in Ref~\cite{paper}.}.
%\begin{small}
%\bea
%P_{\mu \tau} = \sin^{2}(2\theta_{23})\sin^2\left(\frac{\Delta_{31} L}{2}\right)\nonumber\\
%-2|\eta_{\mu \tau}|\sin\delta_{\mu \tau}\sin(2\theta_{23})\sin\left(\Delta_{31} L\right)
%+4|\eta_{\mu \tau}|^2.
%\label{probmutau}
%\eea
%\end{small}
%\noindent
This equation indicates that the CP-odd interference term is only suppressed linearly in $|\eta_{\mu \tau}|$.
\begin{figure}[t]
\vspace{-0.0cm}
\begin{center}
\begin{tabular}{cc}
\hspace{-0.55cm} 
                 \includegraphics[width=7.5cm]{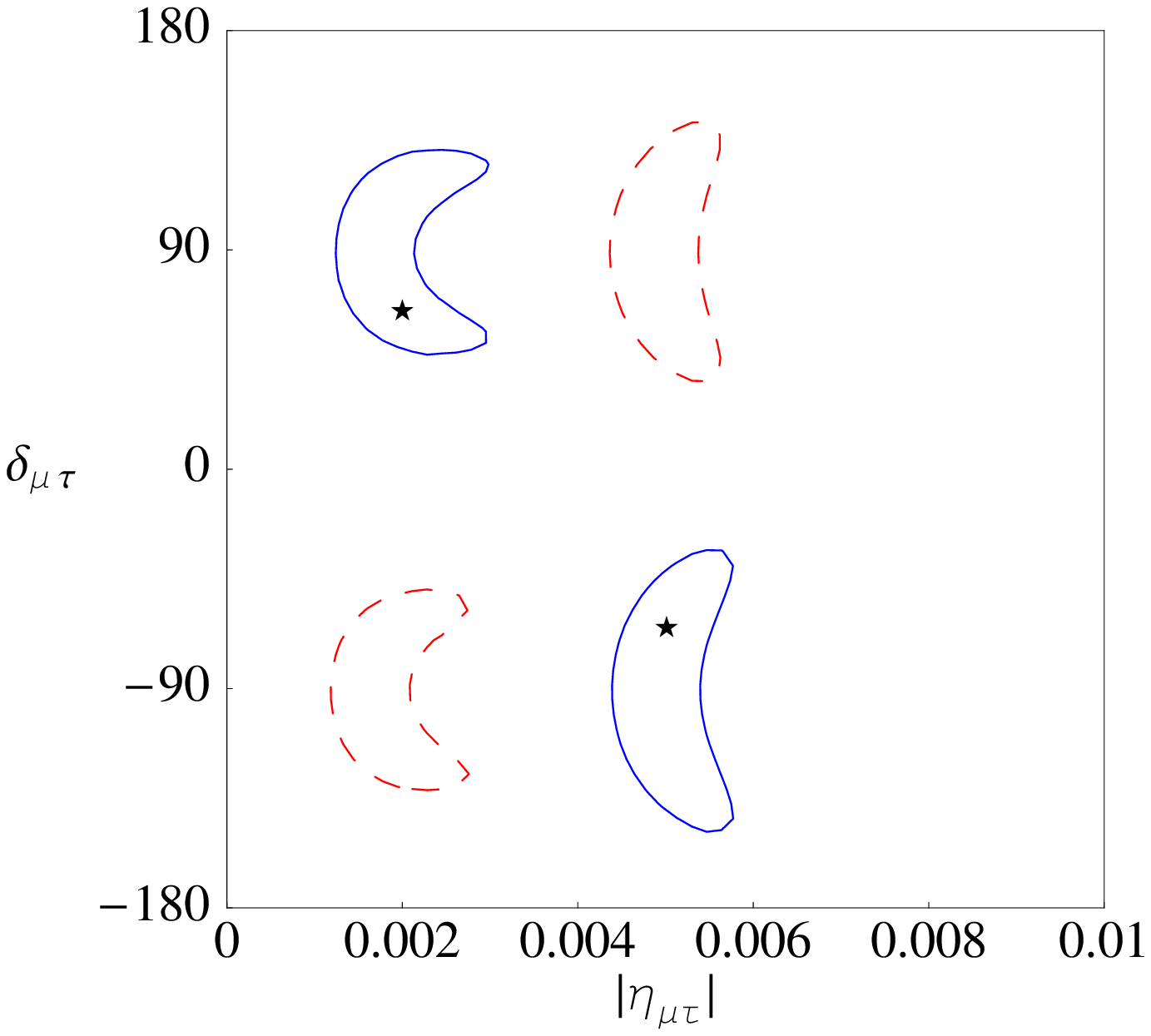} &
		 \includegraphics[width=7.5cm]{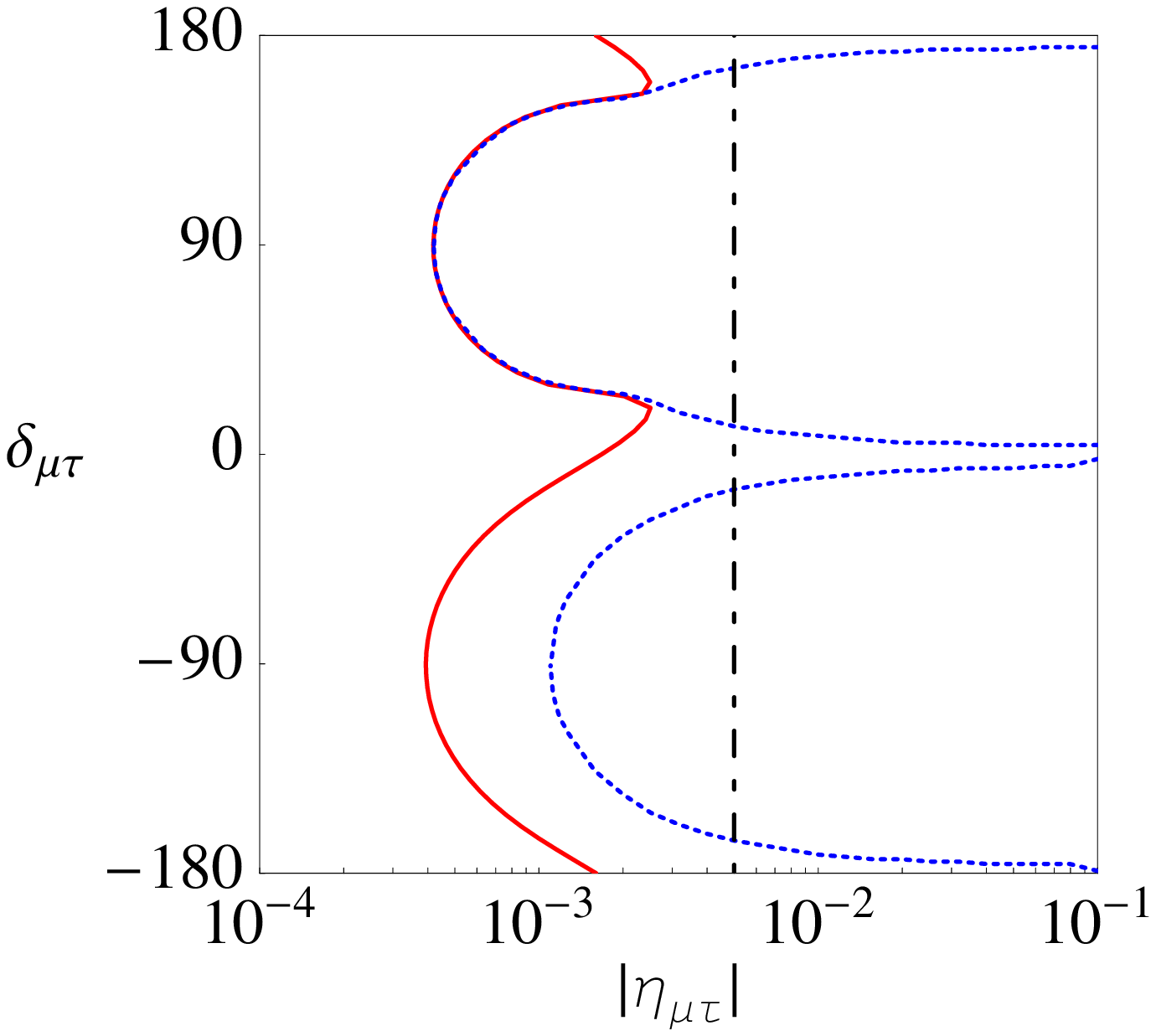}
\end{tabular}
\caption{\it Left: $3 \sigma$ contours for two input values of $|\eta_{\mu \tau}|$ and $\delta_{\mu \tau}$ represented by
the stars. Right: the
solid line represents the $3\sigma$ sensitivity to $|\eta_{\mu \tau}|$ as a function of $\delta_{\mu \tau}$, the dotted
line the $3\sigma$ sensitivity to $\delta_{\mu \tau}$ and the dotted-dashed line represents
the present bound from $\tau \raw \mu \gamma$.
}
\label{fig:mutau}
\end{center}
\end{figure}
This can indeed be observed in the result of the complete numerical computation, Fig.~\ref{fig:mutau}, which shows the sensitivities to $|\eta_{\mu \tau}|$ and $\delta_{\mu \tau}$ obtained. The left
panel represents two fits to two different input values of $|\eta_{\mu \tau}|$ and $\delta_{\mu \tau}$ (depicted by stars). The dashed lines correspond to fits done assuming the {\it wrong} hierarchy, that is the opposite sign for $\Delta_{31}$ to that with which the number of events were generated. As expected from Eq.~(\ref{prob}), a change of sign for the mass difference can be traded by a change of sign
for $\delta_{\mu \tau}$. Nevertheless, this does not spoil the potential for the discovery of CP violation, since a non-trivial value for  $|\delta_{\mu \tau}|$ is enough to indicate CP violation.
 Furthermore, the sinusoidal dependence implies as well a degeneracy between $\delta_{\mu \tau} \raw 180^\circ-\delta_{\mu \tau}$, as reflected  in  the figure.

The right panel in Fig.~\ref{fig:mutau} depicts the $3\sigma$ sensitivities to $|\eta_{\mu \tau}|$ (solid line) and $\delta_{\mu \tau}$ (dotted line), while the present bound from $\tau \raw \mu \gamma$ is also shown (dashed line). The poorest sensitivity to $|\eta_{\mu \tau}|$,
around $10^{-3}$, is found in the vicinity of $\delta_{\mu \tau} = 0$ and $\delta_{\mu \tau} = 180^\circ$, where the CP-odd interference term
 vanishes and the bound is placed through the subleading $|\eta_{\mu \tau}|^2$ term. The latter is also present at zero distance and its effects were already considered in Ref.~\cite{uni}, obtaining a bound of similar magnitude.
 The sensitivity to $|\eta_{\mu \tau}|$ peaks around $|\eta_{\mu \tau}| \simeq 4\cdot 10^{-4}$ for
$\delta_{\mu \tau} \simeq \pm 90^\circ$, where $\sin\delta_{\mu \tau}$ is maximum. That is, for non-trivial values of $\delta_{\mu \tau}$
not only CP-violation could be discovered, but values of $|\eta_{\mu \tau}|$ an order of magnitude smaller could be probed.

\section{Conclusions}
\label{conclusions}

The flavour mixing matrix present in leptonic weak currents may be generically non-unitary, as a result of new physics responsible for neutrino masses. Even if the effects are expected to be extremely tiny in the simplest models of neutrino masses, it is important to analyze the low-energy data without assuming a unitary leptonic mixing matrix, since it is a generic window of new physics. 

A non-unitary matrix has more moduli and phases than a unitary one. The last ones may lead to new signals of CP-violation. In particular, we have shown that an asymmetry between the strength of  $\nu_\mu\rightarrow \nu_\tau$ oscillations versus that for $\bar\nu_\mu\rightarrow \bar \nu_\tau$ results to be a beautiful and excellent probe of new physics, considering short-baselines ($\sim 100$ km.) and a  Neutrino Factory beam of energy $\mathcal{O}(20 GeV)$. Non-trivial values of the new phases can also allow to probe the absolute value of the moduli down to $10^{-4}$, an improvement of an order of magnitude over previous analyzes of future facilities.

\section*{Acknowledgments}
We acknowledge illuminating discussions with Florian Bonnet and Adolfo Guarino during a recent brief stay in Paris. We are also indebted for stimulating comments to Beatriz Ca\~nadas and F. Jimenez-Alburquerque. %Furthermore, the author received partial support of Ministerio de Educaci\'on y Ciencia (MEC) through the project FPA2006-05423, as well as from the Comunidad Aut\'onoma de Madrid through Proyecto HEPHACOS; P-ESP-00346 and of the European Union throug the networking activity BENE, RII3-CT-2003-506395. 
This work was partially supported by the European Union through the "Marie Curie" grant, the author also acknowledges financial support from the MEC through the FPU grant with ref. AP2005-1185.

\end{document}